\documentclass[preprintnumbers,nofootinbib,noshowpacs,eqsecnum,prd]{revtex4}

\usepackage{graphicx}
\usepackage{amsmath,amsthm,amssymb}
\usepackage{mathrsfs,bbm}
\usepackage{array,color}
\usepackage{slashed}
\usepackage{tabularx}

\makeatletter       

\renewcommand{\p@subsection}{}

\makeatother

\begin{document}
\title{Application of BCFW-recursion relations and the Feynman-tree theorem\\ to the four gluon amplitude with all plus helicities}

\author{M. Maniatis}
    \email[E-mail: ]{maniatis8@gmail.com}
\affiliation{Centro de Ciencias Exactas, 
UBB, Casilla 447, Chill\'a{}n, Chile}


\begin{abstract}
Recently it has been shown that in gauge theories amplitudes to any perturbation order
can be obtained by glueing together simple three-point on-shell amplitudes. These three-point 
amplitudes in turn are fixed by locality and Lorentz invariance. 
This factorization into 
three-point on-shell amplitudes follows
from the BCFW recursion relations 
and the Feynman-tree theorem. In an explicit example, that is, 
the four-gluon amplitude with all plus helicities, we illustrate the method.
In conventional calculation this amplitude corresponds to 
one-loop box diagrams.
\end{abstract}

\maketitle

\section{Introduction}
A milestone in the understanding of scattering amplitudes are the 
BCFW recursion relations \cite{Britto:2004ap,Britto:2005fq}: by analytic continuation of the external momenta, 
 tree amplitudes factorize into elementary building blocks of three-point 
amplitudes in any gauge theory or in gravity.
All external particles as well as inner lines are kept on-shell and gauge-invariance is
respected by all subdiagrams. 
An impressive example are the $n$-gluon scattering tree amplitudes which require in calculations
in conventional Feynman diagrams for instance for $n=5$ the computation of 
25 diagrams, for $n=6$ the number of diagrams increases to 220. 
In the maximal helicity violation (MHV) case, that is, 
with two external gluons $i$ and $j$ carrying minus helicity and all others plus helicity, 
the amplitude is notably simple,
\begin{equation} \label{Parke}
A_n(1^+, \ldots, i^-, \ldots, j^-, \ldots, n^+) =
\frac{\langle i  j \rangle}
{\langle 1 2 \rangle \langle 2 3 \rangle \ldots \langle n 1 \rangle}, 
\end{equation}
 as first conjectured by 
 Parke and Taylor  \cite{Parke:1986gb}. 
 With the help of the BCFW recursion relations the amplitude \eqref{Parke} for arbitrary $n$ can be easily proven by complete induction; see for instance \cite{Elvang:2013cua}. 
However, even that the BCFW recursion relations are very powerful, they are limited to tree diagrams
which of course form an unphysical subset of diagrams to a certain perturbation order.\\

A lot of effort has been spent 
to extend the recursion relations beyond tree level, especially to the one loop order:
it has been shown that any amplitude at the one-loop order of a gauge theory can be 
written as a sum over a finite basis of scalar integrals $I_m^{(i)}$ with rational coefficients $c_m^{(i)}$ 
only depending on the outer momenta,
with $m=2,3,\ldots,D$, and $D$ denoting the number of dimensions
\cite{vanNeerven:1983vr,Bern:1994zx,Bidder:2005ri,Anastasiou:2006jv,Giele:2008ve},
\begin{equation} \label{scbasis}
A^{\text{1-loop}} =
\sum_i c_D^{(i)} I_D^{(i)}
+
\sum_j c_{D-1}^{(j)} I_{D-1}^{(j)}
+
\ldots
+
\sum_k c_{2}^{(k)} I_{2}^{(k)}
+
{\cal R}\;,
\end{equation}
and ${\cal R}$ denoting a rational term of the kinematic variables. 
Obviously, the rational term does not have any branch cuts. In the one-loop case a basis of scalar integrals is known. Applying cuts on both sides of \eqref{scbasis} the left-hand side represents products of tree amplitudes and can be calculated. On the right-hand side the cut contributions can be determined from the known expressions for the scalar integrals. In particular,  only those scalar integrals contribute which have the appropriate propagators.  
In this way the coefficients $c_m^{(i)}$ can be determined, that is, together with the known scalar integrals, the amplitude.
The cuts considered are a generalization of the unitary Cutkosky cuts  \cite{Cutkosky:1960sp}, 
which correspond to Feynman diagrams cut into two separate parts.
Generalized cuts denote diagrams where
all possible propagator lines are cut, not necessarily 
splitting the diagram into two parts.
 Unitary cuts provide the 
discontinuities of a diagram but since the rational part in \eqref{scbasis} of the amplitude has no discontinuities in four space-time dimensions, this part is not accessible by unitary cuts.
However, in ${\cal N}=1$ and  ${\cal N}=4$ super Yang Mills (SYM) models it has been shown that also the rational parts can be deduced by unitary cuts \cite{Bern:1994cg,Bern:1994zx,Britto:2004nc,Bidder:2005ri}. This comes from the fact that in these supersymmetric theories the rational parts appear always together with logarithms and polylogarithms carrying discontinuities.
Another interesting observation is that also in non-supersymmetric theories the rational part  
can be determined by treating the momenta not in 4, but in general $D$ dimensions \cite{Bern:1995db,Bern:1996ja,Brandhuber:2005jw}.


The generalized cut methods have been applied, for instance, to the one-loop multi-gluon amplitudes \cite{Bern:2005hs,Bern:1993qk,Mahlon:1993si}. Let us mention in this context the application of single cuts in $D$ dimensions, calculating the all plus helicity four and five gluon amplitudes at one loop order \cite{NigelGlover:2008ur}. New approaches to perform the double cuts have been revealed in \cite{Mastrolia:2009dr} based on Stokes' theorem.
In \cite{Brandhuber:2005kd} it has been shown that in SYM theories one-loop amplitudes with arbitrary helicities of the external particles can be derived from the corresponding maximally helicity-violating amplitudes. 
With respect to single cuts new techniques have been developed in \cite{Britto:2010um}.\\

R. Feynman has shown decades ago that loops may be opened recursively by the application of generalized cuts 
\cite{Feynman:1963ax, Feynman:FTT}. Opening the loops means that propagators are 
replaced by on-shell pairs of particles antiparticles
in the forward limit. The loop integrations are turned into phase-space integrations. 
The Feynman-tree theorem recursively opens all loops: 
any loop is expressed in terms of generalized cut diagrams which reduce the loop order about at least one unit. Iterative application opens all loops and expresses the original Feynman diagram as a sum of tree diagrams. \\

Many new aspects of the Feynman-tree theorem in multi-loop diagrams have been revealed  \cite{Catani:2008xa,CaronHuot:2010zt, Bierenbaum:2010cy,Baadsgaard:2015twa}. 
In particular, it has been shown that only a subset of possible generalized cuts contribute in multi-loop diagrams.
\\

Recently it has been shown \cite{Maniatis:2015kex,Maniatis:2016nmc,Maniatis:2016gui} that the application of the 
Feynman-tree theorem 
followed by the BCFW recursion allows to factorize amplitudes at any loop order in terms of elementary three-point amplitudes in gauge theories. 
In general, an $n$-loop Feynman diagram is turned into a set of tree diagrams after $n$ iteration steps of the Feynman-tree theorem. 
Eventually, arriving at a form with all loops opened, the tree amplitudes can be factorized by the BCFW recursion relations. 
In particular, the method is not limited to a certain perturbation order and gauge invariance is respected by all subamplitudes. 

The application of the Feynman-tree theorem followed by the BCFW recursion relations can be reversed and the amplitudes be constructed by glueing together elementary on-shell three-point amplitudes. Following the BCFW recursion relations, the outer momenta have to be deformed, that is, analytically continued, in order to keep internal lines on-shell without violating momentum conservation. 
To a certain perturbation order, all possible 
tree-diagrams have to be considered. Following the Feynman-tree theorem, in this process particle-antiparticle pairs in the forward limit have to be taken into account. 
These pairs are unobservable but contribute in general to the corresponding perturbation order. Over the phase space of the unobservable particle pairs in the forward limit has to be integrated. 
The singularities originating from the particles in the forward limit can be regularized 
dimensionally. 

Since the three-point scattering amplitudes follow, apart from a coupling constant, from locality and little-group scaling, this means that scattering amplitudes eventually result from these first principles along with unitarity.  
Moreover, every single contributing amplitude is manifestly gauge invariant. Let us note that in a gauge theory like QCD the four-gluon vertex has not to be considered separately, since it follows automatically from glueing together three-point on-shell amplitudes.

We shall illustrate the method in an explicit example, the one-loop four gluon, all plus helicity amplitude. 
We  show how we can express this loop amplitude in terms of three-point on-shell amplitudes glued together. 
This amplitude is an excellent frame to study the methods, since it is rather simple but reveals the main steps of the calculation.


\section{Glueing together on-shell subamplitudes}

We want to consider the all plus helicity four gluon amplitude $A_4(1^+, 2^+, 3^+, 4^+)$.
In conventional Feynman diagram calculation to lowest order this amplitude
follows from a one-loop diagram. For a complex scalar ($s$) circulating in the loop, 
this amplitude reads \cite{Bern:1991aq, Bern:1995db}, 
\begin{equation} \label{AFFeyn}
A_{4}^{s, \text{1-loop}} (1^+, 2^+, 3^+, 4^+) = 
\frac{i}{(4 \pi)^{2-\epsilon}}
\frac{[12] [34]}{\langle 12 \rangle \langle 34 \rangle}
I_4
\end{equation}
with
\begin{equation}
\begin{split}
I_4
= &
- i (4\pi)^{2-\epsilon} 
\int \frac{d^{-2 \epsilon} \mu}{(2\pi)^{-2\epsilon}} \mu^4
\int \frac{ d^4 l}{ (2\pi)^4 }
\frac{1}
{ (l_1^2-\mu^2)(l_2^2-\mu^2)(l_3^2-\mu^2)(l_4^2-\mu^2) }\\
= &  - \frac{1}{6} + {\cal O}(\epsilon).
\end{split}
\end{equation}
The propagator momenta in the loop are given by 
\begin{equation} \label{lmom}
l_1=l, \quad l_2=l-p_1, \quad l_3=l-p_1-p_2, \quad l_4=l-p_1-p_2-p_3=l+p_4 .
\end{equation}
With respect to the Weyl spinors we follow the notation of the review \cite{Elvang:2013cua}.
We shall now show how to decompose this 1-loop 
four-point Feynman diagram into basic building blocks of $A_3$ on-shell subamplitudes. 
The inversion of this procedure is then exactly the process of glueing together 
elementary $A_3$ on-shell amplitudes. 

First, let us note that the all plus helicity tree amplitude for $n$ gluons vanishes,
\begin{equation} \label{ngluontree}
A_{n} (1^+, 2^+, \ldots, n^+) = 0  .
\end{equation}
This can be seen for instance from the factorization via the BCFW-recursion relations 
eventually into three-point amplitudes of the kind $A_3(1^+, 2^+, 3^+)$ which vanish since all gluons carry plus helicities.  
Due to momentum conservation the only non-vanishing three-point on-shell amplitudes are the maximally helicity violating (MHV) or anti-MHV amplitudes with complex momenta.

Therefore, non-vanishing four-gluon amplitudes can only arise beyond tree level. Further, since 
there is no counter term available, which would be proportional to the tree amplitude, there can 
not appear divergences at the one-loop order. The logarithms, respectively polylogarithms 
appear with the singularities. Therefore, we expect the one-loop amplitude  
to be a rational expression in the kinematic variables. 

Applying supersymmetric Ward identities it has been shown \cite{Bern:1996ja} 
that in ${\cal N}=4$ and ${\cal N}=1$  SYM, the relation \eqref{ngluontree} not only holds at tree level, but to all orders in perturbation theory. 
For instance, in ${\cal N}=4$ SYM the 
one-loop gluon amplitude consists in one contribution with a  gluon ($g$, spin 1) 
in the loop, besides four Weyl fermions ($f$, spin 1/2), as well as three complex
scalars ($s$, spin 0),
\begin{equation}
A^{{\cal N}=4, \text{1-loop}}_{n} = 
A^{g, \text{1-loop}}_{n} +
4 A^{f, \text{1-loop}}_{n} +
3 A^{s, \text{1-loop}}_{n}.
\end{equation}
In contrast, in ${\cal N}=1$ SYM we have in the loop to consider one gluon paired with one Weyl fermion,
\begin{equation}
A^{{\cal N}=1, \text{1-loop}}_{n} = 
A^{g, \text{1-loop}}_{n} +
A^{f, \text{1-loop}}_{n}.
\end{equation}
Since the four-gluon amplitudes vanish in ${\cal N}=4$ and  ${\cal N}=1$ SYM, that is, $A^{{\cal N}=4, \text{1-loop}}_{4}(1^+, 2^+, 3^+, 4^+) = 0$, respectively
$A^{{\cal N}=1, \text{1-loop}}_{4}(1^+, 2^+, 3^+, 4^+)=0$, we arrive at the relations \cite{Bern:1996ja}
\begin{equation}
A^{g, \text{1-loop}}_{4}(1^+, 2^+, 3^+, 4^+) = 
-A^{f, \text{1-loop}}_{4}(1^+, 2^+, 3^+, 4^+) =
A^{s, \text{1-loop}}_{4}(1^+, 2^+, 3^+, 4^+) .
\end{equation}
Therefore it suffices to focus on the calculation of the amplitude corresponding to a complex scalar ($s$) in the loop. 
Considering this loop diagram we will encounter as 
elementary building blocks the gluon-scalar-scalar on-shell three-point amplitude. 
This amplitude is, apart from a coupling constant,
fixed by little group scaling and from locality. For simplicity we set the coupling constant to one.
Depending on the helicity of the gluon
 these elementary three-point amplitudes read; see for instance
\cite{Badger:2005zh,Arkani-Hamed:2017jhn}
\begin{equation} \label{A3b}
A_3(1, 2^+, 3) =
\frac{ \langle q |1 | 2]} {\langle q\; 2 \rangle}\;,
\qquad
A_3(1, 2^-, 3) =
-\frac{ \langle 2 |1 | q ]} {[q\; 2]}
\end{equation}
with complex momenta 1 and 3 for the scalars and the gluon with momentum 2 with plus helicity, and 
$q$ an arbitrary linearly independent null vector. 
As we will see later with respect to the BCFW recursion relations, it is convenient to choose this vector $q$ to be the shifted momentum of
the opposite side of the factorized amplitude. 

We shall now show how the conventional Feynman diagram can be factorized into the elementary three-point amplitudes \eqref{A3b}.
We start considering the
1-loop box integral as a conventional Feynman diagram and
in a first step we apply the Feynman-tree theorem \cite{Feynman:1963ax, Feynman:FTT}.
The Feynman-tree theorem, valid at any loop order and for arbitrary dimensions $D$, reads, considering a loop of a Feynman diagram with loop momentum $l$,
\begin{equation} \label{FeynmanTT}
0 =
\int \frac{d^D l}{(2\pi)^D}  N(l) \; \prod_i \bigg\{ G_F^{(i)}(l-p_1-\ldots-p_{i-1}) -
2 \pi \; \delta^{(+)}((l-p_1-\ldots-p_{i-1})^2 -\mu_i^2) \bigg\}.
\end{equation}
where $G_F^{(i)}(p)  = \frac{i}{p^2 - \mu_i^2 + i \epsilon}$ is the $i$'th propagator 
in the loop and $N(l)$ is the 
numerator depending on the details of the model. 
From the expansion of the product on the right-hand side of \eqref{FeynmanTT} we see that we 
can express the original amplitude in terms of all different generalized cut diagrams. 
The generalized cuts are defined by the replacements of the propagators by $-2\pi$ times the $\delta^{(+)}$ distributions,
\begin{equation}
\frac{i}{p^2 - \mu^2 + i \epsilon} \to - 2 \pi \;\delta^{(+)}(p^2 - \mu^2) ,
\end{equation}
where, as usual,
$\delta^{(+)}(p) = \theta(p_0) \delta(p)$. The cut, that is, the $\delta^{(+)}$ distribution puts the momentum originating from the propagator on-shell.
Iteratively we can open all the loops and archive a set of cut diagrams which are all tree diagrams. 
Applying the Feynman-tree theorem \eqref{FeynmanTT} to the one-loop four-gluon amplitude \eqref{AFFeyn} we get by one iteration step four single-cut diagrams, six double-cut diagrams,
four triple-cut diagrams and one quadruple-cut diagram. However, since the outer particles are
 on-shell, the quadruple cut and all triple-cuts vanish immediately since they isolate at least one  three-point amplitude with real momenta. 
 From the six
double-cut diagrams only two survive as well as all four single cut diagrams. The double-cut diagrams which do not leave any isolated on-shell, and therefore vanishing, $A_3$ amplitude, are the horizontal and the vertical cut 
of the diagram.
All not immediately vanishing single and double cuts which 
arise from the Feynman-tree theorem are shown in Fig.~\ref{box-cut1}. 

Since we are applying this theorem to a 1-loop Feynman diagram, there is only one recursion step needed in order to get tree diagrams. In general, a $n$-loop Feynman diagram requires $n$ recursion steps in order to open all loops. The number of cut diagrams for a loop with $i$ propagators is $2^i-1$. In general, 
only a subset of these diagrams contributes, respectively, there are simple relations among these diagrams, as we will explicitly see in the amplitude considered here. 
Note, that the double cuts appearing on the right-hand side of Fig.~\ref{box-cut1} correspond to the usual unitarity cuts, where the whole diagram is split into two parts.

We see that all four single-cut diagrams are related 
by changing cyclically $ 1 \to 2 \to 3 \to 4 \to 1$. 
Similar, the two double-cut diagrams, that is unitary cut diagrams, are related by 
exchanging $ 2 \leftrightarrow 4$. This, in terms of the Mandelstam invariants, $s= (p_1 + p_2)^2$ 
and $t = (p_2+p_3)^2$, corresponds to $ s \leftrightarrow t$. 
\newcolumntype{D}{>{\centering\arraybackslash} m{0.1\textwidth} }  
\begin{figure}[htp]
\begin{tabular}{Dm{20pt}Dm{15pt}Dm{10pt}Dm{10pt}Dm{10pt}Dm{15pt}D}
\includegraphics[height=0.2\textwidth]{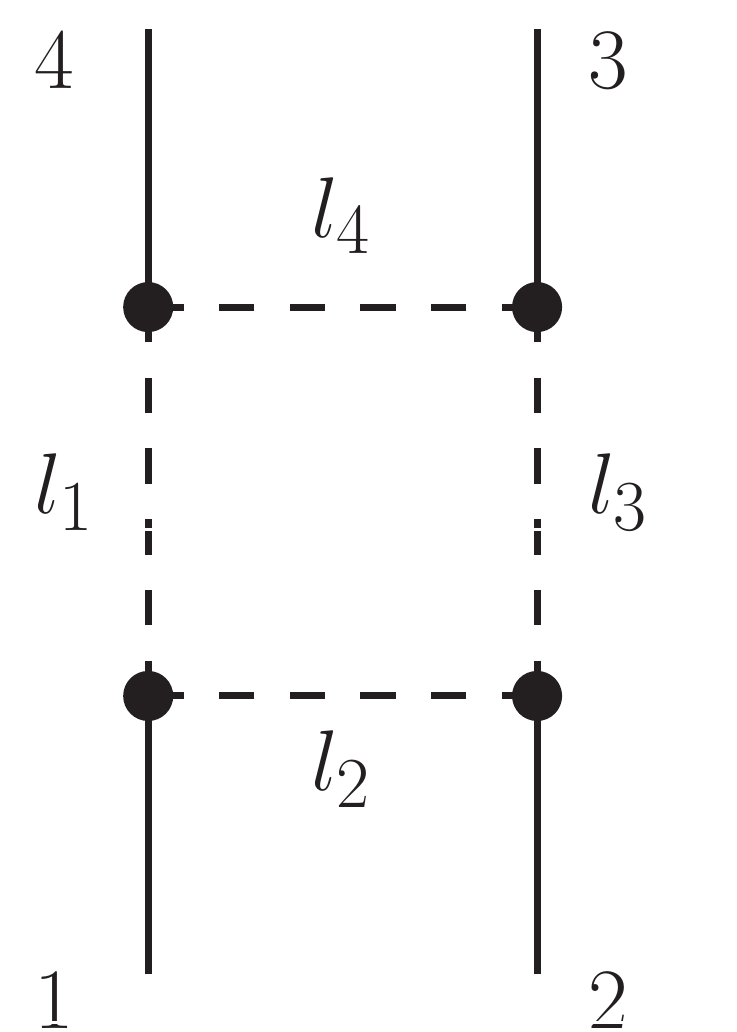}&
\hfill = &
\includegraphics[height=0.2\textwidth]{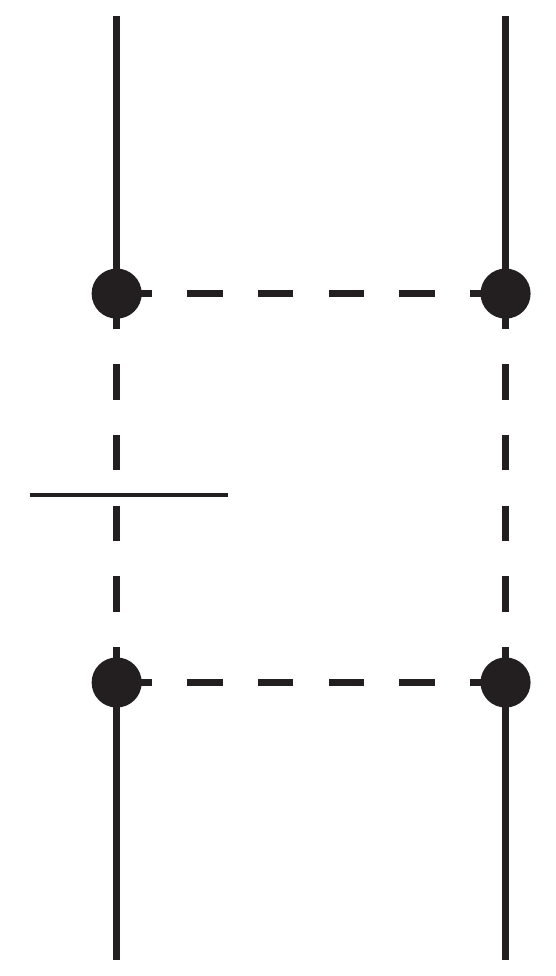} &
\hfill + &
\includegraphics[height=0.2\textwidth]{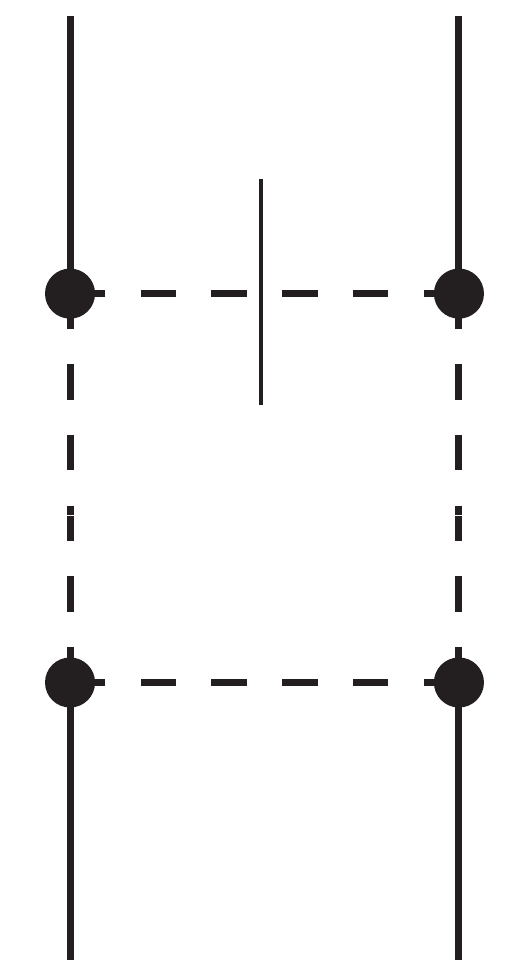}&
\hfill + &
\includegraphics[height=0.2\textwidth]{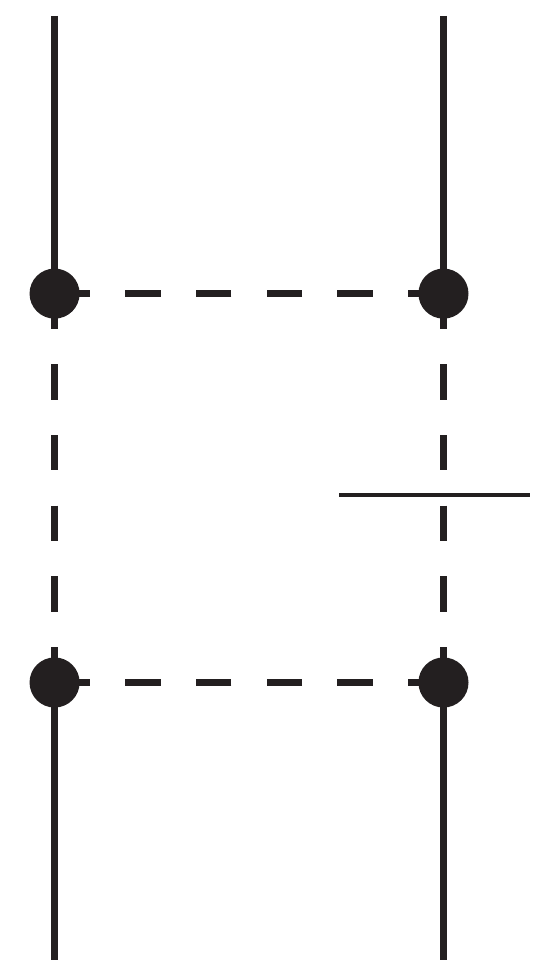} &
\hfill + &
\includegraphics[height=0.2\textwidth]{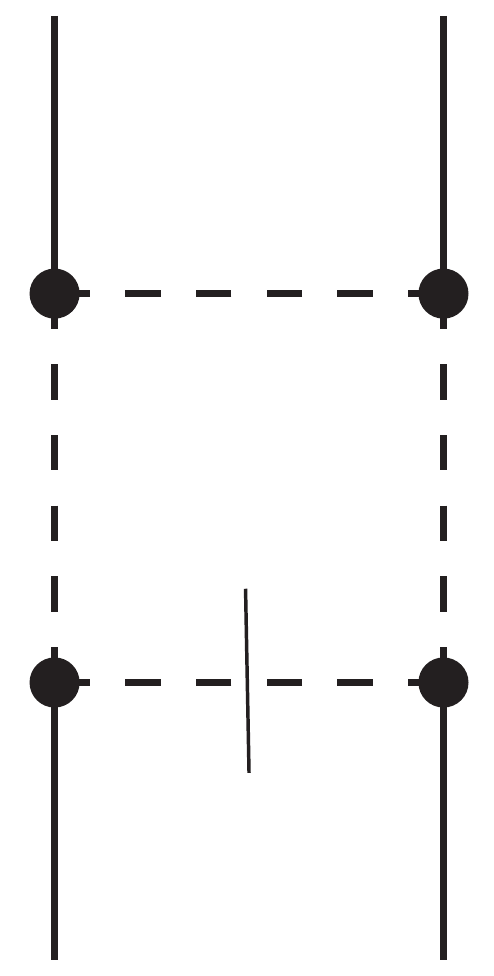} &
\hfill + &
\includegraphics[height=0.2\textwidth]{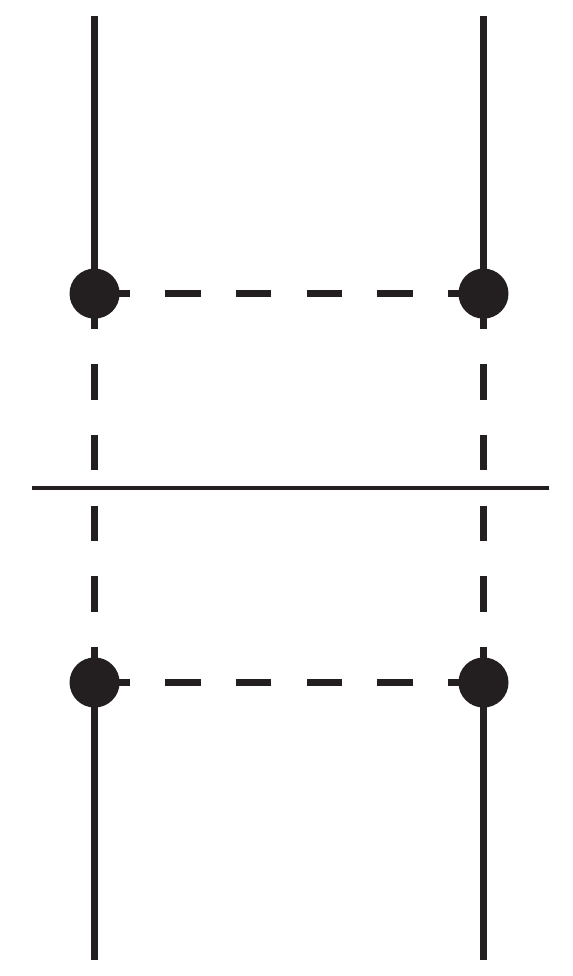} &
\hfill +&
\includegraphics[height=0.2\textwidth]{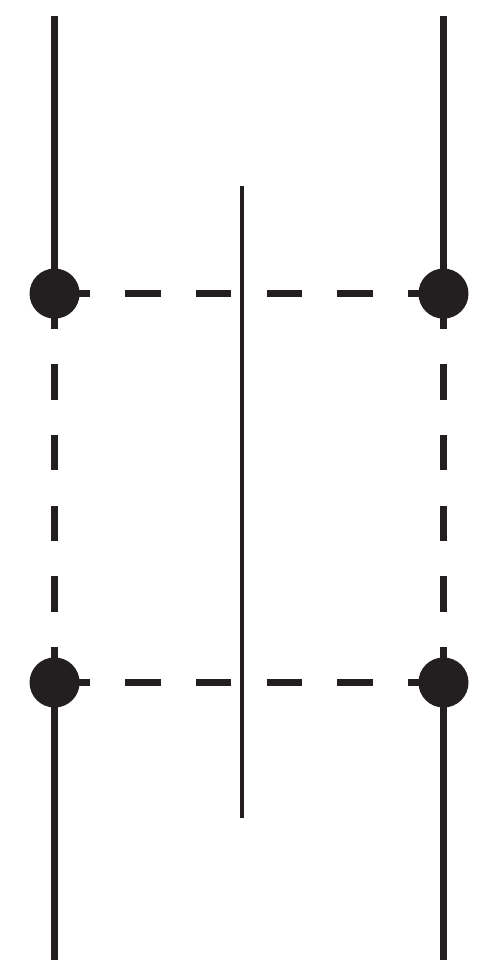}
\end{tabular}
\caption{\label{box-cut1} The  Feynman box diagram and its decomposition 
in terms of generalized cut diagrams. All other cut diagrams vanish because they isolate 
a vanishing on-shell subamplitude $A_3$ with real momenta.}
\end{figure}

We work in the four-dimensional helicity scheme \cite{Bern:1991aq}, that is, 
the external momenta 
are kept four dimensional while the loop momenta are treated in $D$ dimensions.
As has been shown, the $D$-dimensional momentum can be split 
into its four l and $(-2 \epsilon)$ $\mu$ components, that is, $l \to l + \mu$. 
The integration measure can therefore be replaced by
$d^D l \to d^{-2\epsilon} \mu \; d^4 l$
and the propagators in the amplitudes transform as $(l - p)^2 \to (l-p)^2 - \mu^2$.
We get a four dimensional loop integration with an artificial mass parameter $\mu$.\\

Let us start considering  the two double-cut diagrams.  
As we have argued above, from the six double-cut diagrams we 
have four which vanish immediately since they isolate an on-shell three-point amplitude $A_3$, vanishing for real momenta. The remaining two are shown in Fig.˜\ref{box-cut1} on the right-hand side with the indicated unitary cuts, that is, in our case double cuts. As mentioned before, 
the two double-cut diagrams 
are related by exchanges of outer momenta. We calculate therefore only one of them, that is,
the diagram with the momenta $l_1$ and $l_3$ cut as shown in the next-to-last diagram in Fig. \ref{box-cut1}  and get
\begin{multline} \label{A4FTTu}
A_{4, \text{cut} 13}^{\text{1-loop}} (1^+, 2^+, 3^+, 4^+) =\\
\int \frac{d^{-2 \epsilon} \mu}{(2\pi)^{-2\epsilon}}
\int \frac{ d^4 l}{ (2\pi)^4 }
(-2\pi)\delta^{(+)}(l_1^2-\mu^2)
(-2\pi)\delta^{(+)}(l_3^2-\mu^2)
A_{4}(-l_1, 1^+, 2^+,l_3)
A_{4}(-l_3, 3^+, 4^+,l_1) .
\end{multline}
Since we want to decompose the amplitude into its elementary $A_3$ building blocks it remains
to factorize  the four-point  amplitudes which appear in the integrand.
In fact, these amplitudes are tree amplitudes and therefore we can apply the BCFW recursion relations.
We consider a general four point-amplitude of two gluons and two scalars, where the two gluons carry plus helicities, that is,
$A_4(1, 2^+, 3^+, 4)$. We apply a
$[2,3\rangle$ shift yielding with the propagator momentum $P=p_1+p_2$,
\begin{equation} \label{A4}
\begin{split}
A_4(1,2^+, 3^+,4) =  &
A_3(\hat{2}^+, 1, -\hat{P}) \frac{1}{P^2 - \mu^2} A_3(\hat{P}, \hat{3}^+,4)\\
= &
- 
\frac{\langle \hat{3} | \hat{P} | \hat{2} ] } {\langle \hat{3} \hat{2} \rangle}
 \frac{1}{P^2 - \mu^2}
\frac{\langle \hat{2} | 4 | \hat{3} ] } {\langle \hat{2} \hat{3} \rangle}
= 
\frac{\mu^2 [23]}
{\langle 23 \rangle (P^2 -\mu^2) } ,
\end{split}
\end{equation}
where we used cyclicity of the amplitudes and have plugged in the elementary 3-point amplitude \eqref{A3b} with convenient reference vectors.

We  insert the factorization \eqref{A4} into
\eqref{A4FTTu} 
and get 
\begin{equation} \label{A4FTTu2}
A_{4, \text{cut} 13}^{\text{1-loop}} (1^+, 2^+, 3^+, 4^+) =
\frac{ [12][34]}{\langle 12 \rangle \langle 34 \rangle}
\int \frac{d^{-2 \epsilon} \mu}{(2\pi)^{-2\epsilon}}
\mu^4
\int \frac{ d^4 l}{ (2\pi)^4 } 
\frac{
(-2\pi)\delta^{(+)}(l_1^2-\mu^2)
(-2\pi)\delta^{(+)}(l_3^2-\mu^2)
}
{(l_2^2-\mu^2)(l_4^2-\mu^2)} .
\end{equation}
Writing the loop integral in terms of a double integral with a 4-dimensional delta distribution,
\begin{equation} \label{double}
\int \frac{ d^4 l}{ (2\pi)^4 } =
\int \frac{ d^4 l_1}{ (2\pi)^4 }  \frac{ d^4 l_3}{ (2\pi)^4 }  (2\pi)^4 \delta^{(4)}(p_1+p_2-l_1-l_3)
\end{equation}
and performing the integrations over the energy components of $l_1$ and $l_3$ with
the help of the $\delta^{(+)}$ distributions, this amplitude
corresponds to a phase-space integral over two unobservable pairs of particles. 
On the other hand, from the 
Cutkosky cutting rules this phase space integral equals twice the imaginary part of the 
corresponding box integral. However, we know that 
the imaginary parts can only originate from discontinuities of logarithms and polylogarithms.  
Since the scalar box integral has to be a rational function we conclude that 
this integral vanishes.  

The $t$ channel double cut follows from exchanging momenta $2 \leftrightarrow 4$. 
Considering the kinematic factor we find from momentum conservation
$[12]= \langle 41 \rangle [12]/ \langle 41 \rangle = - \langle 4 | 1 | 2]/ \langle 41 \rangle =
 \langle 4 | 3 | 2]/ \langle 41 \rangle = - \langle 43 \rangle [32]/ \langle 41 \rangle$
and similar we have $[34]=- \langle 21 \rangle [14]/ \langle 23 \rangle$, that is, 
we have cyclic invariance of the kinematic factor,
\begin{equation} \label{kinfac}
 \frac{ [1 2] [3 4] }{ \langle 1 2 \rangle \langle 3 4 \rangle}
 =
 \frac{ [2 3] [4 1] }{ \langle 2 3 \rangle \langle 4 1 \rangle} \;.
\end{equation}
From this symmetry we find that the two double-cut integrals are equal, that is, both vanish.\\

\begin{figure}[htp]
\newcolumntype{C}{>{\centering\arraybackslash} m{0.2\textwidth} }  
\begin{tabular}{C c m{5pt} C c}
\includegraphics[width=0.3\textwidth]{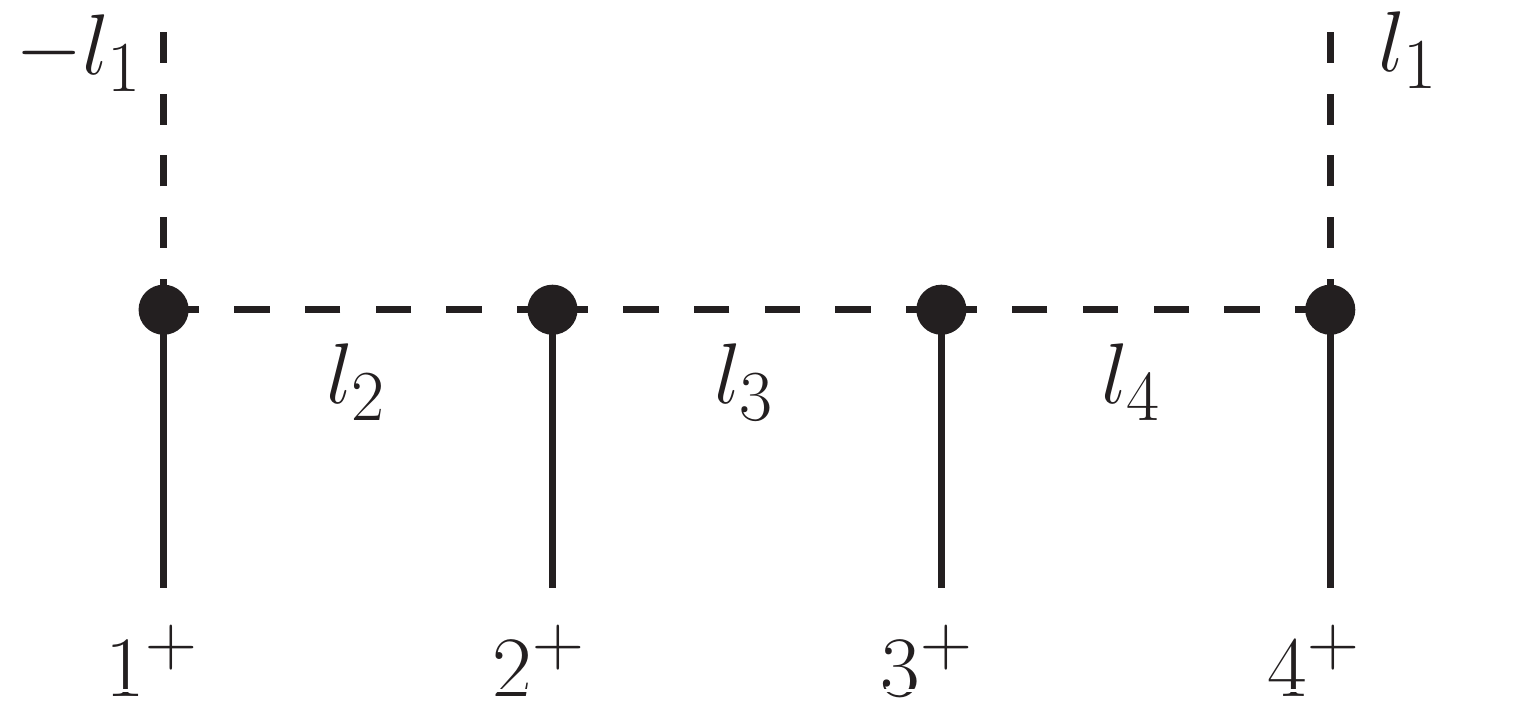}  
& 
\qquad \qquad  \qquad&
= 
& 
\includegraphics[width=0.3\textwidth]{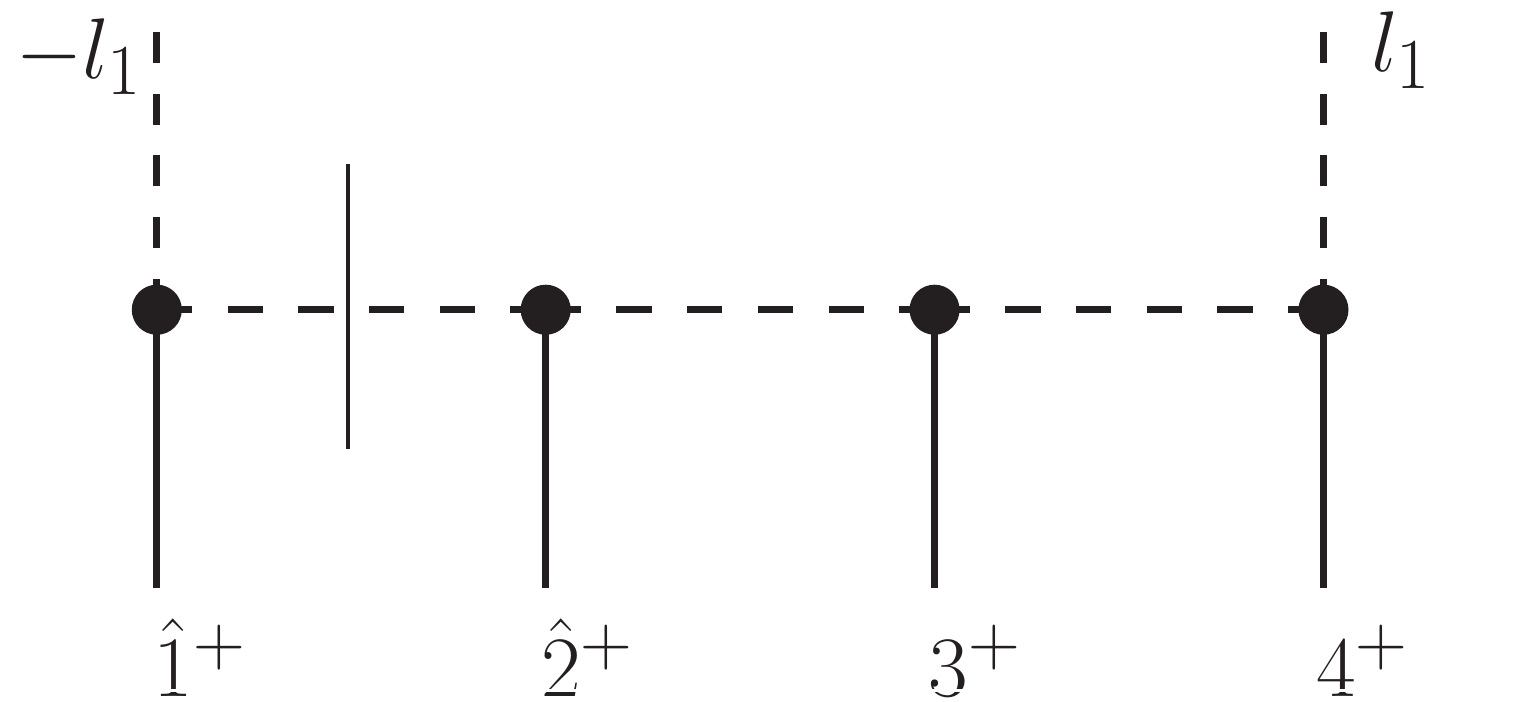} 
& 
\qquad \qquad \qquad \qquad \qquad \qquad
\end{tabular}
\caption{\label{box-1cutBCFW} The single cut of the box diagram and its factorization.
For a $[1,2\rangle$ shift all other cut diagrams vanish because they leave 
both shifted momenta on one side of the propagator.}
\end{figure}

We  now want to compute the single-cut diagram with the cut applied to
the propagator $l_1$  between the gluons 1 and 4 as shown in the first term on the right-hand side of the equation  in Fig.~\ref{box-cut1}. All other single-cut diagrams follow from
cyclic permutations of the outer momenta. The diagram with a single cut of the $l_1$ propagator reads
\begin{equation}  \label{A4single}
A_{4, \text{cut}1}^{\text{1-loop}} (1^+, 2^+, 3^+, 4^+) =
\int \frac{d^{-2 \epsilon} \mu}{(2\pi)^{-2\epsilon}}
\int \frac{ d^4 l}{ (2\pi)^4 }
(-2\pi)\delta^{(+)}(l_1^2-\mu^2)
A_{6}(-l_1, 1^+, 2^+, 3^+, 4^+,l_1) .
\end{equation}
The tree amplitude in the integrand is a six-point amplitude with one unobservable particle-antiparticle pair of complex scalars besides four gluons. The two complex scalars are in the forward limit. 
 We proceed decomposing this six-point amplitude applying the BCFW recursion relations. Since the propagators appear with the mass parameter $\mu$ we 
have to apply the recursion relations adopted to the massive case \cite{Badger:2005zh}. 
We follow the calculation performed in \cite{Badger:2005zh}, where we here have the
case of the two complex scalars in the forward limit.
Applying the BCFW recursion relations, we have to sum over all possible factorizations, corresponding 
to each propagator in turn on-shell. We choose a  $[1,2\rangle$ shift with only one non-vanishing 
contribution as shown in Fig.~\ref{box-1cutBCFW}. All other
single cuts leave both shifted momenta on one side of the BCFW factorization and therefore vanish.
We get
\begin{equation} \label{A4FTTm}
A_{6}(-l_1, 1^+, 2^+, 3^+, 4^+,l_1) =
A_{3}(-l_1, \hat{1}^+, \hat{l}_2)
\frac{1}{l_2^2-\mu^2}
A_{5}(-\hat{l}_2, \hat{2}^+, 3^+, 4^+, l_1) .
\end{equation}
The $[1,2\rangle$ shift  reads explicitly
\begin{equation}
|\hat{1}] = |1] + z |2], \quad
|\hat{2}] = |2], \quad
|\hat{2}\rangle = |2\rangle - z |1\rangle, \quad
|\hat{1}\rangle = |1\rangle ,
\end{equation}
where the complex number $z$ is given by 
\begin{equation}
z= \frac{l_2^2 -\mu^2}{\langle 2 | l_2 | 1 ]} .
\end{equation}

One of the amplitudes coming from the BCFW recursion \eqref{A4FTTm} is already an
elementary three-point amplitude $A_3$. 
In a second recursion step we factorize the five-point amplitude $A_5(1, 2^+, 3^+, 4^+, 5)$, where we choose
conveniently again a shift of the two neighboring gluons, that is, a $|2,3\rangle$ shift:
\begin{equation} \label{A4FTT5}
A_{5}(1, 2^+, 3^+, 4^+, 5)
 =
 A_{3}(1, \hat{2}^+, \hat{P}) 
 \frac{1}{P^2-\mu^2}
A_{4}(-\hat{P}, \hat{3}^+,4^+,5) .
\end{equation}
With another BCFW iteration step to factorize the four-point amplitude with two complex scalars \eqref{A4} into three-point amplitudes we get 
\begin{equation} 
\begin{split}
A_{5}(1, 2^+, 3^+, 4^+, 5)
= & \mu^2
\frac{\langle \hat{3} | 1| \hat{2} ]}{\langle  \hat{3} \hat{2} \rangle}
\frac{1}{P^2-\mu^2}
\frac{1}{(p_4+p_5)^2-\mu^2}
\frac{ [\hat{3} 4]} {\langle \hat{3} 4 \rangle} \\
= &
\frac{\mu^2}
{ [(p_1 + p_2)^2 - \mu^2] [(p_5+p_4)^2-\mu^2]}
\frac{ [4 2] \langle 2 1 \rangle [1 2] + [4 3] \langle 3 1 \rangle [1 2] }
{\langle 2 3 \rangle \langle 3 4 \rangle}
\end{split}
\end{equation}
and in turn (see also \cite{Badger:2005zh})
\begin{equation} 
A_{6}(-l_1, 1^+, 2^+, 3^+, 4^+,l_1)
= 
\frac{\mu^4}
{ (l_1^2- \mu^2)(l_2^2- \mu^2)(l_3^2- \mu^2)}
\frac{ [1 2] [3 4]  }
{\langle 1 2 \rangle \langle 3 4 \rangle} .
\end{equation}
We now plug this result into \eqref{A4single} and eventually get
\begin{equation} \label{A4res}
A_{4, \text{cut} 1}^{\text{1-loop}} (1^+, 2^+, 3^+, 4^+) =
\frac{ [1 2] [3 4] }{ \langle 1 2 \rangle \langle 3 4 \rangle}
\int \frac{d^{-2 \epsilon} \mu}{(2\pi)^{-2\epsilon}}
\mu^4
\int \frac{ d^4 l}{ (2\pi)^4 }
\frac{(-2\pi)\delta^{(+)}(l_1^2-\mu^2) }{(l_2^2-\mu^2) (l_3^2-\mu^2) (l_4^2-\mu^2)} .
\end{equation}
It remains to calculate the integral, where we conveniently go back to $D$ dimensions,
\begin{equation}
I = 
\int \frac{ d^D l}{ (2\pi)^D }
 \frac{-2\pi \delta^{(+)}(l_1^2)}{(l_2^2+i\epsilon) (l_3^2+i\epsilon) (l_4^2+i\epsilon)} .
\end{equation}
We compute this integral with the help of
Schwinger parameters (see also \cite{Baadsgaard:2015twa} for a similar computation), 
\begin{equation} \label{Schwinger}
\frac{i}{x+i\epsilon} \to \int_0^\infty d a e^{iax}, \qquad
 2\pi \delta(x) \to \int_{-\infty}^\infty d a e^{iax}, 
\end{equation}
and get, by integrating over the common Schwinger parameter,  
over the parameter corresponding to the delta distribution, and over
the momentum $l$ in $D$ dimensions
\begin{equation}
I=
\frac{ -i\pi^\frac{D}{2}}{2 (2\pi)^{D}} \Gamma\left(2 - \frac{D}{2}\right)
\int_0^\infty d \alpha_2  d \alpha_3 d \alpha_4 \; [
s \; \alpha_3(\alpha_2 + \alpha_3+ \alpha_4 -1) - t \alpha_2 \alpha_4 
-i\epsilon]^{\frac{D}{2}-2}
\end{equation}
with $s = 2p_1 p_2$ and $t= 2 p_2 p_3$.
With the substitution $\alpha_3 \to \alpha_3 \alpha_4$, integration over
$\alpha_2$, followed by $\alpha_4$, expansion about $D= 4$ dimensions, and
integrating eventually over $\alpha_3$ we get
\begin{equation} \label{res1}
I=
\frac{-i}{32 \pi^2}
\frac{1}{6} 
\left(
	\frac{s}{s+t} + 
	\frac{st}{(s+t)^2} \log\left(\frac{s}{-t}\right) 
\right) + {\cal O}(D-4) .
\end{equation}
The other single-cut diagrams follow from cyclic permutations $1 \to 2 \to 3 \to 4 \to 1$
and we see that we get two equal contributions \eqref{res1} and
two contributions  exchanging $s \leftrightarrow t$. Hence, the kinematic 
factor exactly cancels in the sum and the final result we find for the amplitude is
\begin{equation} \label{result}
A_{4}^s (1^+, 2^+, 3^+, 4^+) =
\frac{-i}{16 \pi^2}
\frac{1}{6} 
\frac{ [1 2] [3 4] }{ \langle 1 2 \rangle \langle 3 4 \rangle}
+ {\cal O}(\epsilon). 
\end{equation}
Applying the Feynman-tree theorem followed by the BCFW recursion relations 
we see that we can reproduce the known result \eqref{AFFeyn}. 


\section{Discussion}

We have seen explicitly that the one-loop box diagram in conventional Feynman diagram calculation 
can be represented in terms of generalized cut diagrams. All triple and quadruple cuts vanish
immediately since they isolate at least one on-shell three-point amplitude with real momenta. Subsequently applying the
BCFW recursion relations, the tree amplitudes factorize into elementary building blocks of three-point amplitudes $A_3$. \\

Reversing the order of the calculation we can construct the amplitude by {\em glueing} together three-point amplitudes. 
The momenta can kept on-shell be their analytic continuation, 
that is, following the BCFW recursion relations, whenever we glue together two on-shell subamplitudes we have
to {\em deform} the outer momenta accordingly. Conveniently this can be done by 
the application of a two-particle shift of two of the external momenta -
one out of each subamplitude. To a certain perturbation order we have to collect all possible constructions of glued on-shell amplitudes. As we have seen in the decomposition, in many cases we encounter vanishing contributions, for instance when we isolate three-point amplitudes with all gluons carrying minus or plus helicities.
To a certain perturbation order we also have to consider amplitudes with pairs of outer particles in the forward limit, that is, particle-antiparticle pairs with opposite momenta and opposite quantum numbers. These pairs are unobservable since they represent vacuum states. Therefore, these
contributions appear in a quite natural way. In the conventional Feynman-diagram approach these amplitudes correspond to loop diagrams. 
Let us emphasize that the Feynman-tree theorem is not limited to the one-loop order, but holds to any loop order. Therefore, 
we can apply the method of glueing together elementary building blocks to any perturbation order. \\

Let us note that in general, the pairs of particles in the forward limit give rise to singularities. However, performing the calculation of the amplitudes in general $D$ dimensions we can regularize these singularities. 

Some remarks to the glueing process are in order: to a certain perturbation order we consider all contributions taking unobservable pairs of particles in the forward limit into account. 
In principle, we may encounter contributions, where we glue together subamplitudes which result in a loop. However this kind of loop is quite different from a loop in the conventional Feynman-diagram approach. In the amplitude approach all outer and inner lines are on-shell in contrast to
Feynman diagrams. 
%
%
%
As we can see, glueing togehter tree diagrams to an on-shell loop diagram we cannot satisfy momentum conservation and on-shellness simultaneously and therefore these contributions vanish naturally. We conclude that by glueing together elementary amplitudes we can disregard any loops. 

\section{conclusion}
The BCFW-recursion relations factorize tree amplitude of gauge theories 
 into elementary three-point on-shell amplitudes which form the elementary building blocks. These elementary building blocks in turn are, apart from a coupling constant, fixed by little group scaling and locality. 
Since we are considering color-ordered amplitudes, the complete amplitudes therefore
 will depend also on the gauge symmetry of the model.\\
 
If we first open the loops by iterative application of the Feynman-tree theorem and then recursively apply the BCFW recursion relations we can factorize any amplitude to any perturbation order into elementary three-point amplitudes. Reversing the recursion relations, we can construct amplitudes by glueing together on-shell three-point amplitudes. To a certain perturbation order we have to consider amplitudes with additional pairs of particles-antiparticles with opposite momenta and opposite quantum numbers, that is, particle pairs in the forward limit corresponding to unobservable vacuum states. \\

We have applied this method to the four-gluon amplitude with all gluons carrying plus helicities to leading order. This amplitude corresponds in conventional calculation to one-loop Feynman diagrams with scalars, fermions, and gluons in the loop. By a hidden supersymmetry all loop contributions can be related to the contribution with a complex scalar running in the loop. Glueing together three-point amplitudes $A_3$ to the forth order in the coupling we have to consider up to four additional particle anti-particle pairs. However, all contributions with three or four particle-antiparticle pairs vanish, because they isolate at least one on-shell three-point amplitude with real momenta. We have seen, that in the case of all plus amplitudes, also the diagrams with two pairs in the forward limit vanish. Eventually we have computed the contributions with one particle pair in the forward limit. In an explicit calculation of  these contributions, we have reproduced the known result for the amplitude. 
Eventually let us emphasize that this calculation of a color-ordered amplitude is based only on locality, little group scaling, that is, Lorentz invariance, and unitarity. 

\acknowledgements{We would like to thank D. Di\'az V\'azquez and P. Mastrolia for helpful discussions. The project was supported in part by the UBB project ``Materia Obscura y los bosones de Higgs'' with number DIUBB 193209 1/R.

\end{document}